\documentclass[%
 aip,
rsi,%
 amsmath,amssymb,
 reprint,%
]{revtex4-1}

\usepackage{graphicx}
\usepackage{dcolumn}
\usepackage{bm}
\usepackage{gensymb} 
\usepackage{textcomp}
\usepackage{xcolor}
\usepackage{colortbl}
\usepackage{array}
\newcolumntype{P}[1]{>{\centering\arraybackslash}p{#1}}

\newcommand{\at}{\makeatletter @\makeatother}
\begin{document}

\preprint{AIP/123-QED}

\title{Demonstration of a picosecond Bragg switch for hard x-rays in a synchrotron-based pump-probe experiment}

\author{M. Sander}
\affiliation{ 
European Sychrotron Radiation Facility, 71 Avenue des Martyrs 38000 Grenoble, France
}%
\affiliation{ 
Tailored X-ray Products gGmbH, Hamburg, Germany
}%
\author{R. Bauer}
\affiliation{%
Instiut f{\"u}r Nanostruktur- und Festk{\"o}rperphysik, Universit{\"a}t Hamburg, Luruper Chaussee 149, 22761 Hamburg
}
\affiliation{ 
Tailored X-ray Products gGmbH, Hamburg, Germany
}%
\author{V. Kabanova}
\affiliation{ 
European Sychrotron Radiation Facility, 71 Avenue des Martyrs 38000 Grenoble, France
}%

\author{M. Levantino}
\affiliation{ 
European Sychrotron Radiation Facility, 71 Avenue des Martyrs 38000 Grenoble, France
}%
\author{M. Wulff}
\affiliation{ 
European Sychrotron Radiation Facility, 71 Avenue des Martyrs 38000 Grenoble, France
}%
\author{D. Pfuetzenreuter}
\affiliation{%
Leibniz-Institut f{\"u}r Kristallz{\"u}chtung, Max-Born-Str. 2, 12489 Berlin, Germany
}%

\author{J. Schwarzkopf}
\affiliation{%
Leibniz-Institut f{\"u}r Kristallz{\"u}chtung, Max-Born-Str. 2, 12489 Berlin, Germany
}%

\author{P. Gaal}
 \email{pgaal@physnet.uni-hamburg.de}
\affiliation{%
Instiut f{\"u}r Nanostruktur- und Festk{\"o}rperphysik, Universit{\"a}t Hamburg, Luruper Chaussee 149, 22761 Hamburg
}
\affiliation{ 
Tailored X-ray Products gGmbH, Hamburg, Germany
}%

\date{\today}

\begin{abstract}
We report a benchmark experiment that demonstrates shortening of hard x-ray pulses in a synchrotron-based optical pump - x-ray probe measurement. The pulse shortening device is a photoacoustic Bragg switch that reduces the temporal resolution of an incident x-ray pulse to approximately 7.5\,ps. We employ the Bragg switch to monitor propagating sound waves in nanometer-thin epitaxial films. With the experimental data we infer the pulse duration, diffraction efficiency and switching contrast of the device. A detailed efficiency analysis shows, that the switch can deliver up to 10$^{9}$ photons/sec in high-repetition rate synchrotron experiments.
\end{abstract}

\keywords{Ultrafast X-ray Diffraction, Photoacoustics, Synchrotron}
\maketitle

\section{\label{sec:level1}Introduction\protect}

Currently, users in the synchrotron community with interest in x-ray pulses with sub-100~ps duration face a changing landscape of facilities. Conditions for time-resolved experiments have significantly improved by the advent of x-ray Free Electron Lasers (XFELs), which provide ultrashort hard x-ray pulses with unprecedented brilliance \cite{XFEL2017a}. Alternatives for few picosecond  and femtosecond hard x-ray pulses are femtoslicing beamlines at ALS \cite{Schoe2000a} and SLS \cite{Ingo2007a, Beau2007a}, table-top plasma sources\cite{zamp2009a,schi2012a} and the new FemtoMax facility at MAXIV \cite{Enqu2018a}. X-ray pulses with few picosecond duration are generated in 3rd generation storage rings using a low charge filling mode, the so-called low $\alpha$ mode\cite{Jank2013a}. This mode reduces the total photon flux due to the low filling charge and is therefore only offered a few weeks per year. Currently the BESSY II synchrotron in planning an upgrade project (BESSY-VSR\cite{BESSYTDR}) which will provide a permanent improved low $\alpha$ mode after installation of additional RF cavities in the storage ring \cite{DiMitri2018}. 

In parallel to the commissioning of XFELs and alternative short pulse sources, many existing synchrotrons are being updated to $4^{th}$ generation low emittance storage rings\cite{Schr2018a}. While low emittance provides better focusing properties and higher beam coherence, the temporal structure, i.e., pulse duration and pulse repetition rate are significantly less favorable for time-resolved experiments. Opportunities for time-resolved experiments at new diffraction limited synchrotron radiation facilities are intensively discussed within the community.

In this article we present a new photoacoustc Bragg switch that allows to shorten hard x-ray pulses emitted from synchrotron storage rings down to few picoseconds. The idea of switching a synchrotron x-ray pulse with a controlled lattice deformation is almost 50 years old\cite{alla1970}. Since then, several attempts were made that relied on piezoelectric excitation\cite{grig2006a,zolo2004a}, generation of optical\cite{buck1999,shep2005} and acoustic\cite{Gaal2014a, Sand2016a} phonons or picosecond thermal \cite{navi2011} excitations. Our device, which we call the PicoSwitch, is tested in a synchrotron-based optical pump - x-ray probe experiment to measure propagation of sound waves in epitaxial nanometer thin films. We discuss important quality parameters, e.g., switching contrast and angle- and time-dependent diffraction efficiency to determine absolute pulse duration and photon efficiency of the shortened pulse. Based on our experimental results we show that the switch can be operated at repetition rates of up to 1\,MHz and delivers pulses of 5-10\,ps duration. The device accepts a limited relative bandwidth of up to $\Delta E/E_0=0.2\%$. At the ID09 beamline at the European Synchrotron (ESRF), where our benchmark experiment was performed, the PicoSwitch can deliver a total flux of up to $10^{9}$ photons/sec, which is the typical intensity from a bending magnet beamline at ESRF. In the following we give a brief introduction of the working principle of the device, which we call the PicoSwitch. A more comprehensive description can be found elsewhere\cite{Sand2016a}.

\begin{figure}
  \centering
  \includegraphics[width = 86mm]{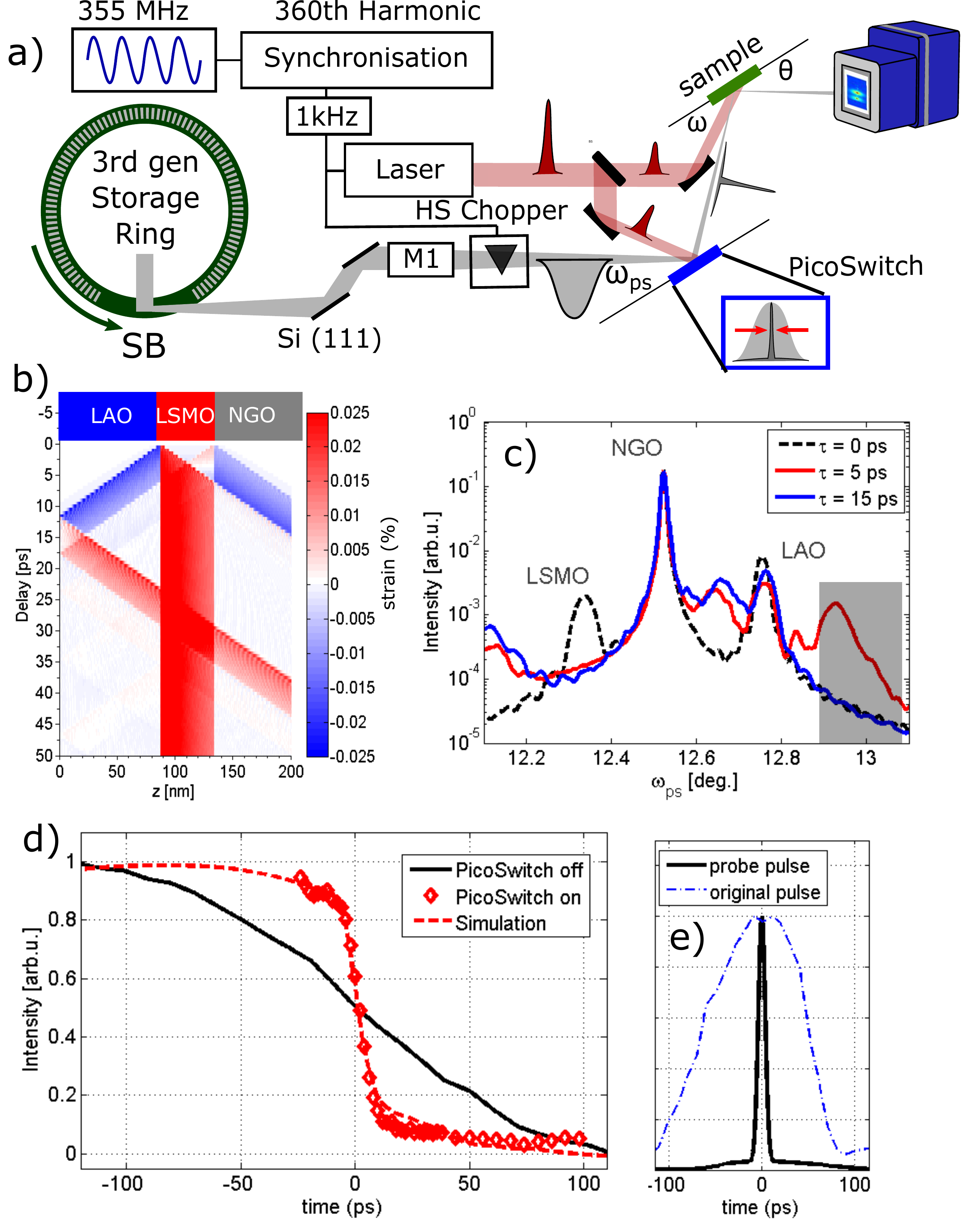}
  \caption{\textbf{Working principle of the PicoSwitch:} a) Experimental setup at the ID09 beamline at ESRF. Two optical pulses from a 1\,kHz amplified laser system are employed to excite the PicoSwitch (blue) and the sample (green). X-ray pulses from the storage ring impinge on the PicoSwitch at an incidence angle $\omega_{ps}$. A shortened x-ray pulse is diffracted to the sample at an incidence angle $\omega$. b) Sketch of the PicoSwitch and sample structure (details in the main text). The plot below shows a spatiotemporal strain map of propagating compression (red) and expansion (blue) waves in the bilayer structures. c) Transient XRD curves of the PicoSwitch after laser excitation calculated from the strain map in b). The simulation for 0\,ps (black dashed line), 5\,ps (red solid line) and 15\,ps (blue solid line) show that the transient diffraction efficiency is turned on and off in the gray shaded area within few picoseconds. d) Pump-probe measurement with the shortened pulse (red symbols) and with the original synchrotron pulse (black solid line) of the ultrafast decrease of diffraction efficiency from a nanostructure. A simulation of the short pulse experiment is shown in the red dashed line. e) Original (blue dashed) and shortened (black) x-ray probe pulse. The original pulse was measured with a correlation technique\cite{gaal2012a}, the shortened pulse was extracted from simulations \cite{Gaal2014a, Sand2016a}}
  \label{fig:setup}
\end{figure}

The layout of the pulse shortening benchmark experiment is shown in Figure~\ref{fig:setup}~a). Unlike conventional pump-probe experiments we employ two optical pump beams from the same laser source, one to trigger the PicoSwitch, the other to excite a sample. The time delay of the first excitation is selected such that the PicoSwitch diffracts incident photons while the maximum intensity of the long synchrotron x-ray pulse is present. Thus, only a temporal slice of the maximum intensity is diffracted from the PicoSwitch, while other parts of the x-ray pulse are suppressed. The shortened x-ray pulse impinges on the sample and is employed to probe dynamics induced by the second optical pump pulse. Since both the PicoSwitch and the sample are excited by optical pulses that stem from the same laser source, the time delay between optical pump and shortened x-ray probe pulse is completely jitter free. The relative pump-probe delay between the optical excitation of the sample and the shortened x-ray pulse is controlled by a motorized delay stage to record the transient sample dynamics up to a pump-probe delay of 2\,ns.

The PicoSwitch structure is shown in the top part of Figure~\ref{fig:setup}~b) and consists of two thin films grown by pulsed laser deposition (PLD) on a dieelectric substrate \cite{SELLMANN2014}. The top layer is composed of a transparent dielectric. The bottom layer is an opaque metal that acts as a thermoelastic transducer upon optical excitation. We underline that several material combinations may be used to build a PicoSwitch device. Here, we use a combination of LaAlO$_3$ (LAO, 85\,nm, transparent) and La$_{0.66}$Sr$_{0.33}$MnO$_3$ (LSMO, 57\,nm, metallic) grown on a NdGaO$_3$ (NGO) substrate. Sound waves generated upon absorption of an optical pump pulse are shown in the colorplot in Figure~\ref{fig:setup}~b) as expansive (red) and compressive (blue) strain. The strain pulses are launched from the interface of the transducer to the adjacent top layer and substrate, respectively.

We calculate the impact of laser-generated strain waves on the angular diffraction efficiency of the PicoSwitch. Results are shown in Figure~\ref{fig:setup}~c) for pump-probe delays of 0\,ps, 5\,ps, and 15\,ps\cite{SCHICK2014}. The transient strain propagating through the PicoSwitch shifts the diffraction efficiency of the top LAO layer to larger diffraction angles and back to the initial position Within 15\,ps. The angular region is marked in gray in Figure~\ref{fig:setup}~c). Here, the PicoSwitch acts as a switchable mirror that is turned on and off by an acoustic pulse on a picosecond timescale.

An important quality parameter of the PicoSwitch is the switching contrast, which describes the suppression of incident x-ray photons during the off-state of the switch. The diffraction efficiency is at a low level $\eta_{0}$ before the arrival of the optical pump pulse. Upon laser excitation the diffraction efficiency mounts to the high level $\eta_{on}$ for a switching time $\Delta T$, which is determined by the propagation of strain waves in the structure. After the coherent strain waves have propagated away from the thin films into the substrate, the diffraction efficiency falls back to its initial value $\eta_{\infty}\approx\eta_{on}$. We define the the switching contrast with the following expression\cite{gaal2012a} 
\begin{equation}
    c_{sw}=\frac{\eta-\eta_{0,\infty}}{\eta_{0,\infty}}
\end{equation}
$c_{sw}$ basically has the same value before and after switching is turned on and off, respectively. This is a significant improvement compared to earlier designs of the PicoSwitch, where the final contrast $c_{sw,\infty}$ was strongly reduced by laser heating of the structure\cite{Gaal2014a}. Whether introducing the PicoSwitch results in high temporal resolution or not depends on the ratio of the pulse area of the original and the shortened pulse, respectively. Therefore we define the total contrast as the product of the switching contrast and the area loss factor, i.e., the ratio of the normalized pulse area of original and shortened pulse: \begin{equation}
    C_{total}=c_{sw}\cdot ALF
\end{equation}
where
\begin{equation}
    ALF=\frac{\int_{-\infty}^{\infty} dt I_{sw}(t)/I_{sw,m}}{\int_{-\infty}^{\infty} dt \tilde{I}(t)/\tilde{I}_{m}}
\end{equation}

and $I_{sw}(t)$, $I_{sw,max}$, $\tilde{I}(t)$ and $\tilde{I}_{m}$ denote time-dependent and maximum intensity of the shortened and original x-ray pulse, respectively. While $\tilde{I}(t)$ can be easily measured with the PicoSwitch\cite{gaal2012a}, $I_{sw}$ is deduced from model calculations as shown below.

The pulse shortening capability is shown in Figure~\ref{fig:setup}~d) which shows a step-like decrease of the diffraction efficiency upon optical excitation. The sample and physical origin of the sudden intensity change are discussed in the next paragraphs. For now, we highlight the influence of the duration of the probe pulse on the measured dynamics. Figure~\ref{fig:setup}~d) clearly shows a dramatic increase of the temporal resolution in the measurement that employs a PicoSwitch. The temporal shape of the corresponding probe pulse is shown in Figure~\ref{fig:setup}~e). The original, long synchrotron pulse has been measured using a fast sampling method\cite{gaal2012a}, which yields a full width at half maximum (FWHM) pulse duration of 120\,ps. We also clearly recognize the expected asymmetric pulse shape. The shortened pulse was derived from simulations and crosschecked by comparison with experimental data, which is shown in the red dashed line in Figure~\ref{fig:setup}~d).

Now we discuss the experimental capabilities of the PicoSwitch pulse shortening scheme in a real synchrotron-based pump-probe experiment. The sample is composed of a similar structure as the PicoSwitch itself, i.e., a thin film system composed of a transparent dieelectric LAO top layer with a thickness of 104\,nm on a metallic LSMO layer with a thickness of 93\,nm. The stack is grown on a NGO substrate. Note that the film thicknesses of the sample and the PicoSwitch are different, which results in slightly different propagation times of the coherent sound wave through the respective structure.

\begin{figure}
  \centering
  \includegraphics[width = 86mm]{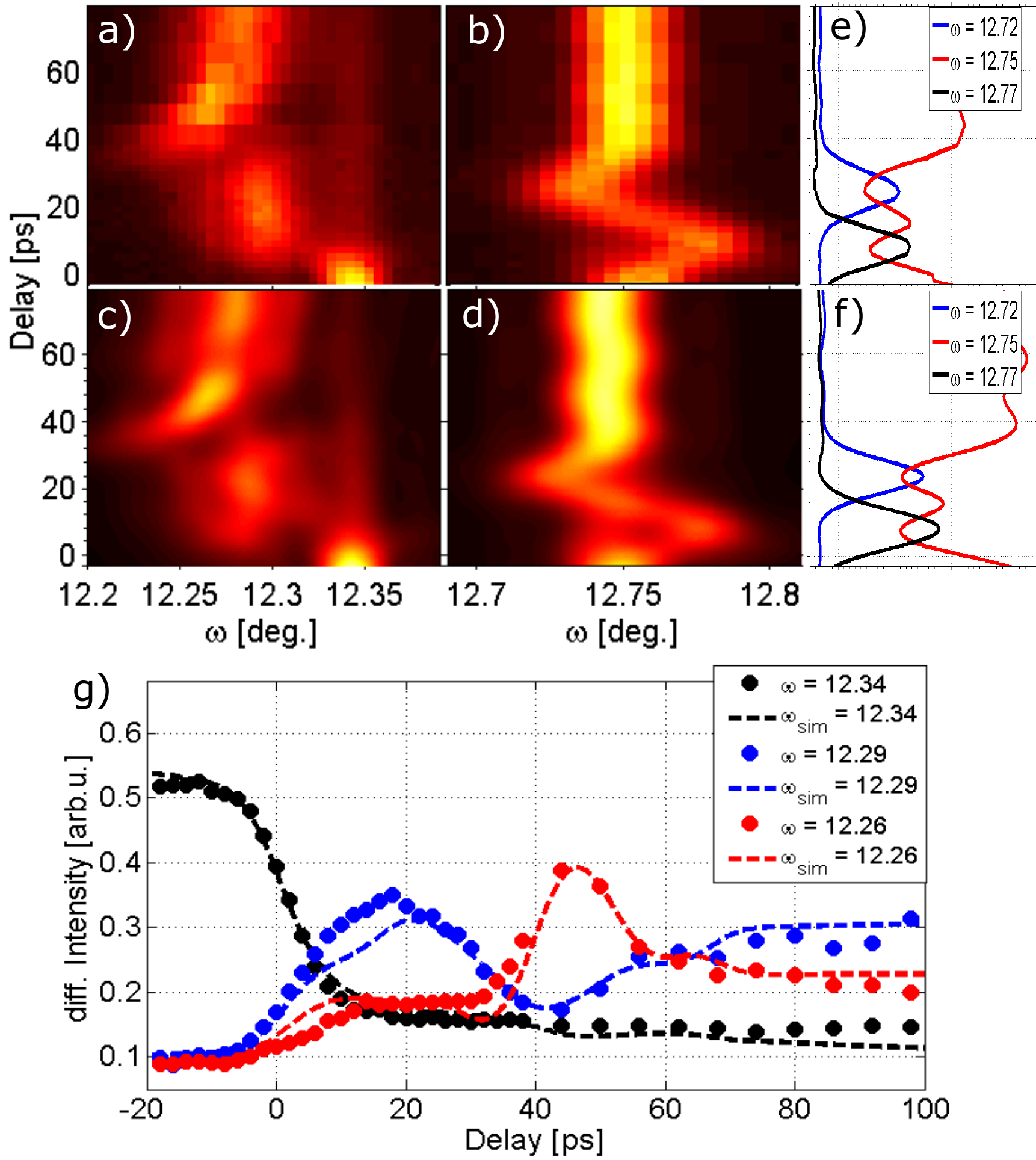}
  \caption{\textbf{Experimental data:} a) and b) Time-resolved measurement of propagating sound waves in the LSMO [a)] and LAO [b)] film of the sample. The measurement employed pulse shortening with the PicoSwitch. Fast picosecond lattice dynamics is clearly resolved by the experiment. The mechanism behind the observed XRD peak shift is outlined in the main text. c) and d) Simulations of propagating sound waves in the LSMO [c)] and LAO [d)] layer. Comparison of simulation and experimental data reveals the x-ray probe pulse duration of 7.5\,ps and the total switching contrast of $C_{total}$=2.94. e)-g) cross sections of measured and simulated data at different incidence angels $\omega$ on the sample.}
  \label{fig:Data}
\end{figure}

The lattice dynamics measured by time-resolved XRD of the metallic LSMO and dielectric LAO layers are shown in Figure~\ref{fig:Data}~a) and b), respectively. Experiments were performed at an x-ray photon energy of 14.85\,keV on the (002) lattice planes in symmetric $\omega$-2$\theta$ geometry. 

Both sample and PicoSwitch were mounted on motorized xyz-translation stages and on a motorized rotation circle with angular resolution better than 0.1\,mrad for tuning the incidence angle of the x-ray beam. The size of the x-ray beam before and after symmetric diffraction from the PicoSwitch was approximately 40x60~$\mu$m$^{2}$. The acceptance angle of the PicoSwitch was 870~$\mu$rad at an angle of 12.9$^{\circ}$ which is 8 times larger than the full vertical divergence in the focus of the x-ray beam. At the sample position, approximately 150\,mm after the PicoSwitch, we observed no changes of the x-ray beam footprint, divergence or stability. In our setup, the shortened beam is deflected upwards. Insertion of a multilayer mirror could be used to deflect the beam downwards, thus yielding a horizontal beam. The repetition rate of the x-ray beam is reduced to 1 kHz by a system of choppers\cite{Camm2009a} to match the laser repetition frequency. The main purpose of the choppers is to reduce the heat load on the beamline optics from the intense x-ray radiation. They also protect soft matter and biological samples from unnecessary radiation damage, thus leading to a longer sample lifetime. It should be noted that the PicoSwitch contrast is not sufficient to gate a single pulse from the synchrotron pulse train. Instead, gating can be performed electronically by modern detectors.\cite{Shay2017a} At ID09 beamline, diffraction and scattering signals are recorded by a Rayonix HS170 detector in accumulation mode, i.e. the signal from many laser/x-ray pulse pairs is accumulated without any time resolution provided by the detector. Beam parameters are summarized in Table~\ref{tab:ESRFparameters}.

\begin{table}
\begin{tabular}{|p{3 cm}|P{2 cm}|P{2 cm}|}
\hline
 & \textbf{ESRF} & \textbf{ESRF-EBS} \\
\hline
\textbf{Focus} & &  \\
\hline
\textbf{H [$\mu m$]} & 40  & 20  \\
\textbf{V [$\mu m$]} & 60  & 20  \\
\hline
\textbf{Divergence} & & \\
\hline
\textbf{H [$\mu rad$]} & 860  & 53 \\
\textbf{V [$\mu rad$]} & 106  & 28 \\
\hline
\textbf{Photons/pulse} & $2.1\cdot10^{9}$ & $3.0\cdot10^{9}$\\
\textbf{(pink beam)} & &  \\
\hline
\textbf{Pulse Duration} & $<$135~ps & $<$150~ps \\
\hline
\end{tabular}
\caption{Main parameters of the ID09 beamline at ESRF. The second column lists the parameters after the EBS upgrade (FWHM values).}
\label{tab:ESRFparameters}
\end{table}

The sample and PicoSwitch were excited with an optical fluence of 30\,mJ/cm$^{2}$. The principal axes of the elliptical laser footprint on the sample and PicoSwitch were 920\,$\mu$m/600\,$\mu$m and 720\,$\mu$m/630\,$\mu$m, respectively.  We performed simulations of coherent acoustic phonon propagation in the sample using a one-dimensional linear chain model of masses and springs \cite{herz2012b}. The phenomenon of propagating high-frequency coherent acoustic phonon wavepackets is well understood \cite{thom1986,rose1999,lars2002,barg2004,boja2013, herz2012c, shay2013a, boja2015}. Our simulations yield a two-dimensional map of lattice strain vs. pump-probe delay along the out-of-plane spatial axis in the sample as shown in Figure~\ref{fig:setup}~b). From the spatiotemporal strain map we calculate transient XRD curves using dynamical diffraction theory \cite{SCHICK2014,warren1990x}. Finally, we convolute the XRD simulation with the simulated x-ray probe pulse after reflection from the PicoSwitch. The FWHM pulse duration of the shortened pulse is approximately 7.5\,ps and the pulse is shown in Figure~\ref{fig:setup}~e). 

Figure~\ref{fig:Data}~c) and d) show simulated dynamics of the LSMO and LAO diffraction peaks, respectively. To reproduce the experimental data in a simulated pump-probe experiment, we find a switching contrast of $c_{sw}$=33. With $ALF$=11.34 we find a total contrast $C_{total}$ of 2.93. Both experiment and simulation show the effect of insufficiently suppressed photons of the original long x-ray pulse, e.g., at the equilibrium angle of the LSMO peak of 12.34$\degree$ in Figure~\ref{fig:Data}~a) and c). Still, the picosecond dynamics of the propagating sound waves is clearly resolved in the measurement. 

Comparing the colorplots shown in Figure~\ref{fig:Data}~a) and c) as well as b) and d) we find excellent agreement of the simulated pump-probe experiment with the experimental data. In particular, we observe a step-like drop of the LSMO peak intensity at $\omega=12.34\degree$ [black dashed line and black bullets Figure~\ref{fig:Data}~g)]. A delay scan at this incidence angle is shown in Figure~\ref{fig:setup}~d), compared to a measurement of the same dynamics with the original ESRF x-ray probe pulse. The LSMO peak reappears at $\omega=12.29\degree$ for approximately 15\,ps [blue dashed line and blue bullets Figure~\ref{fig:Data}~g)]. This new peak position corresponds to the thermal expansion resulting from the energy deposited by the absorbed optical excitation pulse. At a pump-probe delay of $\approx$40\,ps, the LSMO peak is distorted again but returns to its intermediate expanded angular position [red dashed line and red bullets Figure~\ref{fig:Data}~g)]. 

The dynamic features observed at the LSMO peak are well understood by simulations of thermal expansion and coherent phonon generation and propagation\cite{herz2012b}. The initial shift of the LSMO reflex stems from thermal expansion while the second peak distortion originates from the coherent sound wave that is back reflected at the sample surface\cite{Sand2016a}. Having determined the sample geometry by static XRD and ellipsometry measurements, we adjust the simulated dynamics via the sound velocity in LSMO and LAO, respectively, to the experimental data. The result agrees well with values reported by other groups\cite{Bogdanova2003, Michael1992}. For comparison, a similar experiment reported earlier by our group gave 20$\%$ higher sound velocities due to insufficient temporal and angular resolution of our XRD setup\cite{Sand2016a}. 

As depicted in Figure~\ref{fig:Data}~a), we also observe a significant broadening of the LSMO peak after optical excitation. Within the time delay covered in the experiment, thermal transport and heat equilibration does not lead to a significant strain equilibration within the two layers. Hence, due to the high temporal resolution provided by the PicoSwitch and due to the high angular resolution provided by the synchrotron, the initial excitation profile directly after absorption of the laser pulse is resolved. The data yields an exponential decay of the strain in the excited LSMO layer with a decay constant of 55\,nm$^{-1}$.

\begin{figure}
  \centering
  \includegraphics[width = 86mm]{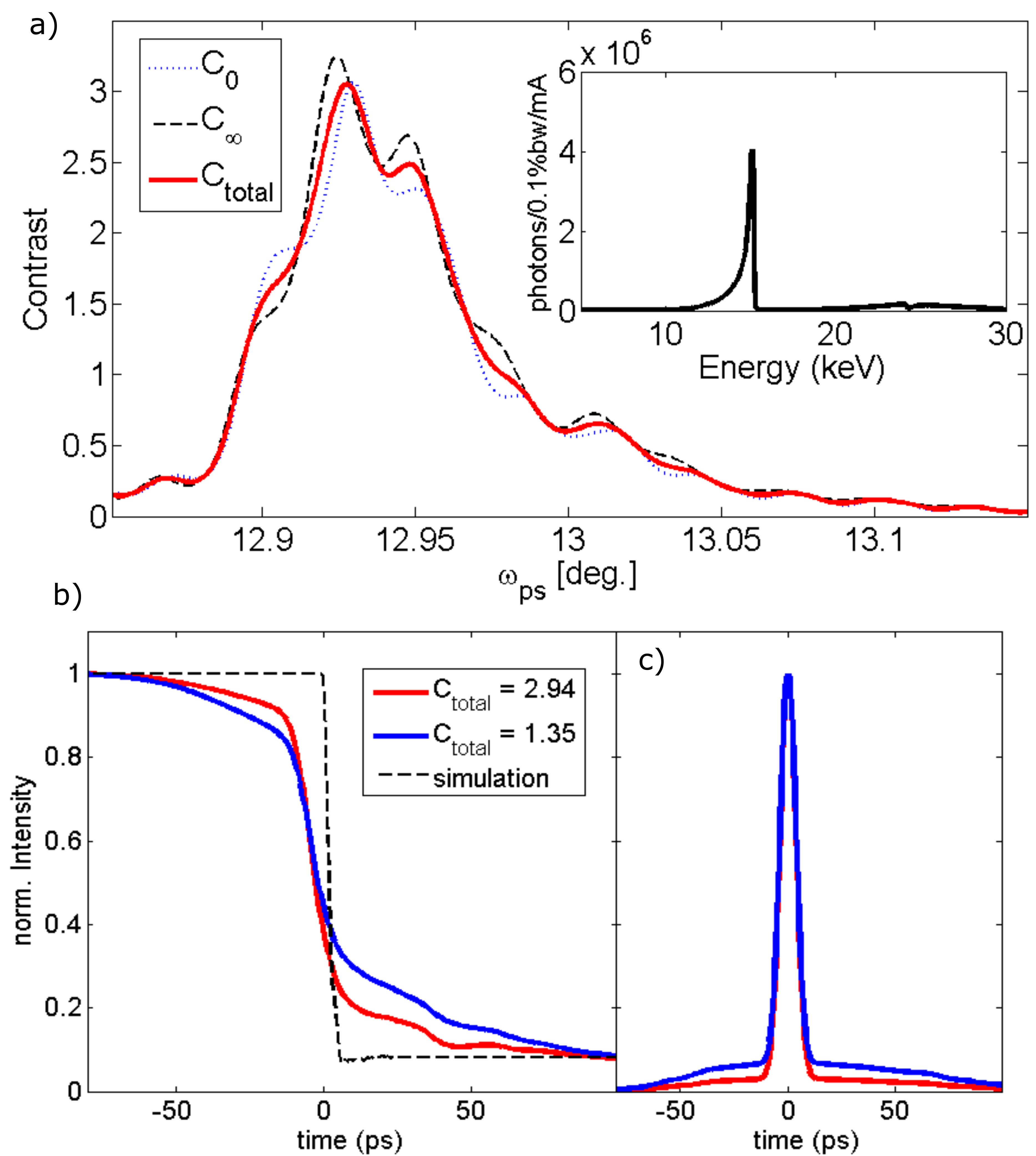}
  \caption{\textbf{Switching contrast:} a) Calculated switching contrast of the PicoSwitch in the gray shaded area of Figure~\ref{fig:setup}~c). The contrast results from the difference in diffraction efficiency in the on and off state of the switch and from the area loss function $ALF = 11.34$. The inset shows the X-ray spectrum emitted by the U17 undulator at ID09 beamline. b) Calculated pump-probe signals of an ultrashort step function (black dashed line) with a contrast of 2.94 (red solid line) and 1.35 (blue solid line), respectively. The latter contrast is obtained at a 10-times higher x-ray bandwidth. d) X-ray probe pulses used for the calculations in c).}
  \label{fig:Contrast}
\end{figure}

Finally, we discuss the photon flux in the shortened x-ray pulse to check the efficiency of the PicoSwitch: by comparing the integral pulse areas of the original and shortened pulses shown in Figure~\ref{fig:setup}~e), we find an Area Loss Factor $ALF$=11.34. The total intensity loss must also account for the finite diffraction $\eta_{on}$ of $2\cdot10^{-3}$. We find a total efficiency of $1.8\cdot10^{-4}$. Our measurement was performed with the U17 undulator at ID09 ESRF which delivers $2.1\cdot10^{6}$\,photons/pulse at an energy of $E_0=14.85$\,keV, a bunch current of 5\,mA and a relative bandwith of $\Delta E/E_0=0.016\%$. With the above considerations, the photon flux reduces to $3.6\cdot10^2$ photons/pulse. The calculated initial, final and total contrast is shown in Figure~\ref{fig:Contrast}~a) for a monochromatic x-ray pulse. The high contrast region also determines a limit of the angular stability of the switch. For experiments which tolerate a higher relative bandwidth the number of photons can be increased almost linearly with the relative bandwidth $\Delta E/E_0$. However, the switching contrast decreases, if the relative bandwidth is too large. The effect of a finite total contrast is demonstrated in Figure~\ref{fig:Contrast}~b) which depicts again the sampling of the LSMO peak (black dashed line) with a probe pulse with 7.5\,ps (FWHM) and a contrast of 2.93 (red solid line) and 1.35 (blue solid line), respectively. The simulated probe pulses are shown in Figure~\ref{fig:Contrast}~c). Clearly increasing the bandwidth of the x-ray probe pulse yields a higher photon flux. However, it goes hand in hand with a decreased switching contrast. Therefore, the total contrast and intensity of the PicoSwitch probe pulse are related quantities, which allow for adaption to specific experiments. We would like to point out again that the PicoSwitch contrast is insufficient for pulse gating. For that, slower photoacoustic transient gratings with diffraction efficiencies of up to 33\% may be used\cite{sand2017a,sand2017b}. Another parameter for optimizing the experiment and the x-ray flux is to increase the repetition frequency of the PicoSwitch. A successful implementation of the PicoSwitch at a repetition rate of 208\,kHz was already presented \cite{Sand2016a} and operation even above 1\,MHz has been tested successfully. Assuming the PicoSwitch is operated at the ESRF orbit frequency of 354\,kHz, the experimentally derived parameters from the measurement shown in Figure~\ref{fig:Data}, yield a total x-ray flux of $1.2\cdot10^8$ photons/sec. The main performance parameters of the short pulse beam are summarized in Tab.~\ref{tab:parameters}. The table also provides values for an optimized beamline setting with increased bandwidth and increased repetition frequency. As discussed above, both parameters increase the x-ray flux in the shortened beam. We also provide the pulse parameters for other synchrotron-based short pulse facilities.

\begin{table}[h]
\centering
\begin{tabular}{|p{2 cm}|P{2 cm}|P{2 cm}|P{2 cm}|}
\hline
\textbf{Parameter}  & \textbf{ID09 (ESRF)} & \textbf{PicoSwitch Experiment} & \textbf{PicoSwitch optimized} \\
\hline
\textbf{Photons/ pulse}  &  &  & \\  
$\frac{\Delta E}{E_{0}}=10^{-4}$ & $2\cdot10^{6}$ & $3.6\cdot10^{2}$ & \\
$\frac{\Delta E}{E_{0}}=10^{-2}$ & $2\cdot10^{8}$ &  & $3.6\cdot10^{4}$ \\
\hline
\textbf{Photons/ sec} & & & \\
\at 1~kHz & $2\cdot10^{9}$ & $3.6\cdot10^{5}$ & \\
\at 100~kHz & $2\cdot10^{11}$ &  & $3.6\cdot10^{9}$ \\
 \hline
\end{tabular}
\begin{tabular}{|P{8.45 cm}|}
\hline
\textbf{Alternative Sources for femto- and picosecond x-ray pulses} \\
\hline
\end{tabular}
\begin{tabular}{|p{2 cm}|P{2 cm}|P{2 cm}|P{2 cm}|}
\hline
\textbf{Parameter}  & \textbf{FemtoMAX\cite{Enqu2018a}} & \textbf{NSLS-II\cite{NSLSII}} & \textbf{BESSY VSR\cite{BESSYTDR}} \\
\hline
\textbf{Photons/ pulse}  & $1\cdot10^{7}$ &  & \\
\hline
\textbf{Energy [keV]}  & 1.8 - 20 & 1 - 20 &  $<$~10 \\
\hline
\textbf{Repetition Rate}  & 100~Hz & 500~MHz & 1.25~MHz  \\
\hline
\textbf{Pulse Duration} & 100~fs & 15~-~30~ps & 15~ps~-~300~fs\\
 \hline
\end{tabular}
\caption{(upper) Performance parameters of the x-ray beam at ID09, the shortened beam during the experiment and an optimized setting with increased bandwidth and repetition rate. For the beam size and beam divergence we also list the beamline parameters after the Extremly Brilliant Source (EBS) upgrade. (lower) Summary of alternative synchrotron based sources for picosecond and sub-picosecond x-ray pulses.}
\label{tab:parameters}
\end{table}

In conclusion, we have demonstrated the feasibility of pulse shortening with fast photoacoustic Bragg switches for synchrotron-based pump-probe experiments. Our device, which we call the PicoSwitch shortens an incident 100\,ps long hard x-ray pulse to a duration of 7.5~ps (FWHM). We have defined and quantified the relevant parameters for the pulse duration, efficiency and switching contrast of the PicoSwitch. Even with the rather low efficiency of a 1~kHz setup, our experiment monitors structural dynamics due to propagating sound waves in thin epitaxial films. In particular, we profit from the superb beam stability and angular resolution of the synchrotron beam, which are not degraded by insertion of the PicoSwitch. In an optimized setup with repetition rates up to 1\,MHz and a bandwidth of the x-ray pusle of 0.2\%, the PicoSwitch would deliver a flux of more than 10$^{9}$~photons/sec. The PicoSwitch is a powerful option for introducing high temporal resolution on a beamline level in synchrotron-based experiments. It may become a valuable tool for time-resolved experiments in current and future large-scale radiation facilities.

We acknowledge financial support from the BMBF via grant 05K16GU3. VK and MW gratefully acknowledge support from the EU grant Horizon 2020 XPROBE (\#637295).


\providecommand{\noopsort}[1]{}\providecommand{\singleletter}[1]{#1}%

\end{document}